\documentclass[reprint, prl, twocolumn]{revtex4-1}

\usepackage{amssymb,amsfonts,amsmath}
\usepackage{mdwlist} 
\usepackage{dcolumn}
\usepackage{graphicx}
\usepackage{epstopdf}
\usepackage{hyperref}
\usepackage{natbib}

\newcommand{\R}{\mathbb{R}}

\newcommand{\eps}{\epsilon}
\newcommand{\bb}[1]{\mathbf{#1}}

\begin{document}

\title{Enumerating rigid sphere packings}

\author{Miranda C. Holmes-Cerfon} 
\affiliation{Courant Institute of Mathematical Sciences, New York University.}

\begin{abstract}
Packing problems, which ask how to arrange a collection of objects in space to meet certain criteria, are important in a great many physical and biological systems, where geometrical arrangements at small scales control behaviour at larger ones. 
In many systems there is no single, optimal packing that dominates, but rather one must understand 
 the entire set of possible packings. 
As a step in this direction we enumerate rigid clusters of identical hard spheres for $n\leq 14$, and clusters with the maximum number of contacts for $n\leq 19$. 
A rigid cluster is one that cannot be continuously deformed while maintaining all contacts. This is a nonlinear notion that arises naturally because such clusters are the metastable states when the spheres interact with a short-range potential, as is the case in many nano- or micro-scale systems.
We expect these lists are nearly complete, except for a small number of highly singular clusters (linearly floppy but nonlinearly rigid.)  
The data contains some major geometrical surprises, such as the prevalence of hypostatic clusters: those with less than the $3n-6$ contacts generically necessary for rigidity. 
We discuss these and several other unusual clusters, whose geometries may shed insight into physical mechanisms, pose mathematical and computational problems, or bring inspiration for designing new materials. 
\end{abstract}

\maketitle

\section{Introduction}

The study of sphere packings has a long and rich history in mathematics \cite{conway1999,saff1997}. 
A large body of work has searched for optimal packings -- for example, those that maximize the density of an infinite collection of spheres in different dimensions \cite{cohn2003,hales2005}, 
or those that minimize an energy or volume functional \cite{tarnai1996,cohn2007,newton2011,wright2013}, such as the Thomson problem which considers electrons on a unit sphere \cite{bowick}. However, many applications call for knowing the \emph{total set} of packings -- all the possible ways to arrange a given, finite number of spheres to satisfy certain conditions. For example, in condensed-matter physics, spheres are used to model atoms, molecules, colloids, or other units of matter, and one asks how large numbers of units behave collectively. 
A rich set of phases can emerge, such as crystals, gels, and glasses, and the dynamics of forming these or changing between them are often controlled by the geometrical ways to arrange small groups of spheres without overlap 
\cite{stillinger1984,nelson1989,frenkel2001,royall2008,meng2010}. Granular materials, such as sand, are also modelled as packings of spheres or other shapes. The total number of mechanically stable packings is argued to give rise to an extensive entropy \cite{edwards1989,frenkel2014}, so once this is known then statistical mechanical ideas may be applied to a system that has so far defied this approach.
In chemistry, the set of clusters of hard spheres is conjectured to have the most ``rugged'' energy landscape, so it may be possible to derive the landscape of molecules with smoother interactions from these \cite{hoare83,wales2001}.
Engineers and materials scientists have proposed to use clusters of nanoparticles as the basis for new materials, but must know the possibilities and how to build them with high yield \cite{manoharan2004,fan2010,schade2013}.

Here we ask: what are all the ways to arrange $n$ spheres  so they form a rigid cluster? A rigid cluster is one that cannot be deformed continuously by any finite amount and still maintain all contacts. We argue this is the most natural set of clusters to consider: first, because rigid clusters are topologically equivalent -- they are all isolated solutions to the system of equations (see \eqref{eq:bonds}) -- and second, because
when the spheres interact with a short-range potential, as is the case for many nano- or micro-scale particles, then these conformations are the local minima on the energy landscape, hence the metastable states where the system spends most of its time \cite{holmescerfon2013}.

The number of clusters is expected to increase very rapidly so we cannot hope to solve this for arbitrary $n$ \cite{stillinger1984,walesdoye1997,frenkel2014}, yet the solution even for small cluster size would provide valuable insight into the problems above. 
In physics, a bar is set at $n=13$. This is the smallest number where one sphere becomes caged: completely surrounded by neighbouring spheres, 
so there is expected to be a richer variety of geometrical structures and physical  implications  \cite{hoare83}. 
Previous work has listed clusters of $n\leq11$ spheres \cite{arkus2011,hoy2012} but these approaches are not easy to extend to higher $n$, and they use a notion of ``rigid'' which is not the most physically natural. 

We attack this question computationally, using a bottom-up, dynamical algorithm to list rigid clusters of $n\leq 14$ spheres, and a subset for $ n \leq 19$ that contains the ground states (clusters with the maximum number of contacts.) 
While we have no proof these sets are complete, there is good empirical evidence to suggest they are missing only a very small number of clusters that are too singular (see below) for the algorithm to handle. 
Thus, for the first time, we produce a nearly complete dataset in a physically relevant size regime.

The data contains several surprises -- clusters that run counter to certain basic physical intuitions or assumptions -- and the main aim of the paper is to highlight these. 
The biggest surprise is the prevalence of  (i) Clusters with \emph{fewer} than $3n-6$ contacts. This runs counter to the commonly-cited Maxwell's criterion, which states that a rigid cluster should have at least $3n-6$ contacts \cite{maxwell1864,roux2000,vanhecke2010,lubensky2012}. 
Other surprises include: 
(ii) Geometrically distinct clusters with the same adjacency matrices  -- i.e. clusters with the same set of contacts but which are not related by rotations, reflections, or permutations; (iii) Ground state clusters that are almost all close-packing fragments (stackings of hexagonal plane fragments) beyond $n\gtrapprox 15$. This occurs at much lower $n$ than for a longer-range potential such as Lennard-Jones; 
 (iv) A sharply decreasing overall proportion of clusters that are close packing fragments;
 (v) A roughly constant proportion of ``singular'' clusters ($\approx$ 2.5\%) -- those that are linearly floppy but nonlinearly rigid; 
 (vi) Clusters with ``circular'' modes of deformation: when a single contact is broken, the cluster continuously deforms until it reforms exactly the same contact in exactly the same configuration;
 (vii) A rigid cluster that has no one-dimensional paths to it: if any contact is broken, it acquires at least two modes of deformation. 

These clusters are intended to stimulate the imagination, but they are more than simply geometrical curiosities. Many of them, such as the singular and hypostatic ones, represent states of matter whose behaviour in a thermodynamic system are not yet understood. They bring data to materials science --- the bulk behaviour of a material arises in part from geometrical arrangements of its most basic components, for which the set of clusters forms a catalogue of all possible motifs. Inventing new materials may be possible by changing the arrangements of a few localized components \cite{paulose2014}, and this catalogue can bring inspiration for unusual geometries that are hard to construct explicitly. 
Additionally, it has been remarked to the author that these clusters have educational value because they are physical representations of unusual solutions to algebraic equations --- ``a nilpotent group you can hold in your hand'' \cite{vakil}. 
While the surprises are not surprising from a mathematical perspective -- indeed, they are solutions to a system of nonlinear equations, so in principle any type of pathology could occur \cite{vakil2006} --- it is valuable for all of the above reasons to have concrete examples where they occur in a physically relevant system. 

A second aim of the paper is to point to problems that could benefit from an applied mathematics perspective. 
This type of packing problem occurs widely, so having a more rigorous understanding of the problem, its physical implications, and the methods used to solve it would be valuable. 

Here is an outline of the paper. In section 2 we precisely define the problem, briefly describe how we test for rigidity, and outline the enumeration algorithm. In section 3 we show results from the algorithm. Section 4 discusses selected aspects of the methods and results. Section 5 is a brief conclusion.

\section{Problem and methods} 

Here is the problem we wish to solve: given $n$ spheres in three dimensional-space with unit diameters, what are all the possible ways to arrange them without overlap so they form a rigid cluster?

Mathematically, let a cluster  be represented as a vector $ \bb{x} = ( \bb{x}_1,  \bb{x}_2, \ldots,  \bb{x}_{n})\in \mathbb{R}^{3n}$ where $ \bb{x}_i = (x_{3i-2}, x_{3i-1}, x_{3i})$ is the center of the $i$th sphere, combined with a set of algebraic equations of the form 
\begin{equation}\label{eq:bonds}
| \bb{x}_i -  \bb{x}_j|^2 - 1 = 0
\end{equation}
 for each pair of spheres $(i,j)$ that are in contact. 
Additionally, we add six equations to remove the translational and rotational degrees of freedom, for example by fixing one sphere at the origin, another on the $x$-axis, and a third on the $xy$-plane, as \begin{equation}\label{eq:cons}
x_{s}=0,  \qquad s\in\{1,2,3,5,6,9\}.
\end{equation} 

The system \eqref{eq:bonds} is often represented by an adjacency matrix $A$ by setting $A_{ij}=1$ if spheres $i,j$ are in contact, and $A_{ij} = 0$ otherwise.  
We consider only clusters where the spheres do not overlap, so require $|\bb{x}_i-\bb{x}_j| \geq 1$ for all $i\neq j$. 

\subsection{Defining and testing rigidity} 

A cluster $ \bb{x}$ with adjacency matrix $A$ is \emph{rigid} if $ \bb{x}$ is an isolated solution to \eqref{eq:bonds},\eqref{eq:cons}, i.e. the solution is a  zero-dimensional point \cite{asimow78,connelly1996,gortler2010}. 
A cluster $ \bb{x}$ is \emph{floppy} if it lies on a positive-dimensional solution set. 

Generically, we expect a rigid cluster to have $3n-6$ contacts. This comes from equating the number of variables ($3n$) to the number of equations, and is the origin of ``Maxwell's criterion'' in the physics literature \cite{maxwell1864}. However, this condition is neither necessary nor sufficient for rigidity.

This notion of rigidity is a nonlinear one (to distinguish it from others, we will sometimes say ``nonlinearly rigid''), and there is a good physical reason to consider it. Suppose the spheres interact with a very short-range potential, as is often the case for mesoscale particles such as colloids \cite{doyewales1996b,meng2010,crocker2011,holmescerfon2013}. Then, the energy of a cluster is basically proportional to the number of contacts, so a floppy cluster can lower its energy by moving along its degrees of freedom until it forms another contact. Once it reaches a rigid cluster it cannot rearrange without breaking a contact to overcome an energy barrier, so these clusters are metastable states where the system spends most of its time in equilibrium. 

Testing for rigidity is a hard problem and there is no known algorithm to do it efficiently. Indeed, determining whether a given configuration of a linkage is rigid is thought to be coNP-hard \cite{demaine2007, abbott2008}. Instead, we test for a stronger condition: that of \emph{pre-stress stability}. This is a concept that comes from structural engineering, and asks that a cluster $\bb{x}$ be the minimum of certain energy function, whose Hessian at the minimum is positive definite \cite{connelly1996,gortler2014}. 

Here is how we test this. Suppose there is a  deformation $ \bb{p}(t)$ depending analytically on parameter $t$, with $ \bb{p}(0) =  \bb{x}$.  Taking one derivative of the system \eqref{eq:bonds},\eqref{eq:cons}  gives  
\begin{equation}
R( \bb{x}) \bb{p}'\vert_{t=0}=0,
\end{equation}
 where $R( \bb{x})$ is the Jacobian of \eqref{eq:bonds},\eqref{eq:cons}, called the \emph{rigidity matrix}. If the right null space of $R( \bb{x})$ is empty, we cannot solve for $ \bb{p}'(0)$ so the cluster is \emph{infinitestimally rigid}, or \emph{first-order rigid}. This is sufficient for rigidity \cite{connelly1996}. 
 
 This is a linear criterion, so we will sometimes say ``linearly rigid'' or ``linearly floppy.'' A cluster that is linearly floppy may or may not be rigid. The right null space of the rigidity matrix gives the linear deformations of the cluster, and to check whether these are extendable to finite deformations we must continue the expansion to higher order.

Taking two derivatives of \eqref{eq:bonds},\eqref{eq:cons} gives  
\begin{equation}
R( \bb{x}) \bb{p}''\vert_{t=0} = -R( \bb{p}') \bb{p}' \vert_{t=0}.
\end{equation}
By the Fredholm alternative, we can solve for $ \bb{p}''(0)$ if and only if there exists $v\in\mathcal{V}$ such that $w^TR(v)v = 0$ for all $w\in\mathcal{W}$, where $\mathcal{V},\mathcal{W}$ are the right and left null spaces of $R( \bb{x})$. When this condition does not hold, the cluster is \emph{second-order rigid} and this is also sufficient for rigidity \cite{connelly1996}.  
Finding a $w$ for each $v$ to make the inner product non-zero is a challenge, but sometimes it is possible to find a single $w$ that works for all $v$. This happens when $w^TR(v)v$ is sign-definite for $v\in\mathcal{V}$, and then the cluster is \emph{pre-stress stable}. 
It is possible to find such a $w$ by semi-definite programming methods, for example. See Supplementary Information for details about how this is implemented. 

To compute the number of internal degrees of freedom of a cluster when it is not pre-stress stable, we use a numerical method that estimates the dimension of the solution set by taking small steps in each of the candidate tangent directions \cite{SI}.

\begin{table*}
{\footnotesize
\begin{center}
\begin{tabular}{c | cccccccc | c}
$n$ & \multicolumn{8}{c|}{number of contacts} \\ 
   & $3n-9$& $3n-8$ & $3n-7$ & $3n-6$ & $3n-5$ & $3n-4$ & $3n-3$ & $3n-2$ &Total \\ \hline
5 & & & & 1 &&&&& 1\\
6 & & & & 2 &&&&& 2\\
7 & & & & 5  &&&&& 5\\
8 & & & & 13  &&&&& 13\\
9 & & & & 52  &&&&& 52\\
10 & & & 1 & 259 & 3  &&&& 263 \\  
11 & & 2 & 18 & 1618 & 20 & 1  & & &  1659 \\
12 & & 11 & 148 & 11,638 & 174 & 8 & 1 & &  11,980 \\
13 & & 87 & 1221 & 95,810 & 1307 & 96 & 8 & &  98,529 \\
14 & 1  &  707  &  10,537  &  872,992  &  10,280 &  878 &  79 & 4 &   895,478 \\\hline
& $3n-4$ & $3n-3$ & $3n-2$ & $3n-1$ & $3n$ & $3n+1$ & $3n+2$ &  &\\
\hline
15 &    7675  & 782  &  55 & 6 & &&&& ($9\times 10^6$ est.)  \\
16 &       &  7895 & 664 & 62 & 8 &&&& ($1\times 10^8$ est.)  \\
17 &      &   &  7796 & 789 & 85 & 6 &&& ($1.2\times 10^{9}$ est.) \\
18  &    &   &   &  9629  & 1085 & 91 & 5 &  &  ($1.6\times 10^{10}$ est.) \\
19  &    &   &   &    &  13,472 & 1458  & 95 & 7 &  ($2.2\times 10^{11}$ est.) \\
\hline
\end{tabular}\\
\end{center}}
\caption{Number of clusters found for each $n$, organized by number of contacts in each cluster. For $n\geq15$ only clusters with a given minimum number of contacts were enumerated. }\label{tbl:results}
\end{table*}

\subsection{Enumeration algorithm}
We search for rigid clusters by following all the one-dimensional transition paths between clusters. We begin with a single rigid cluster of $n$ spheres. This is easy to obtain, for example by gluing a sphere with three contacts to a cluster of $n-1$ spheres. 
Next we break a contact on this cluster, by deleting a single equation in \eqref{eq:bonds}. Typically, this makes a cluster with a single internal degree of freedom, i.e. the set of solutions to the reduced system of equations forms a one-dimensional manifold. When this is the case we follow this one-dimensional path numerically 
until a new contact is formed \cite{SI}. The new contact adds an equation to the system, so the resulting cluster is typically rigid. This cluster could be merely a permutation of the cluster we originally started with, but it could also be an entirely distinct cluster. In the latter case, we add it to our list of rigid clusters. 

For each cluster in our list, we break all subsets of contacts that lead to a one-dimensional transition path, follow each path, and keep track of the clusters at the other end. 
When we reach the end of the list, we stop: we have the entire set of pre-stress stable clusters that are connected to the starting cluster by the one-dimensional paths that are computed.

\section{Results}

This algorithm was run to enumerate rigid clusters of size $n\leq 14$. 
Table \ref{tbl:results} shows the total number of rigid clusters found for each $n$; the coordinates are available in the Supplementary Materials or on the website \cite{mhcwebsite}.  
The total number $N(n)$ increases very quickly: a good fit is  $N(n) \approx 2.5(n-5)!$.
This is \emph{faster} than the exponential increase of clusters with smooth potentials \cite{stillinger1984, walesdoye1997}. The discrepancy may be because the minimum gap between non-contacting spheres in a packing appears to become arbitrarily close to 1 --- for $n=14$ it is 1.3$\times10^{-5}$ \cite{SI} --- but for a smooth potential it is typically bounded away from the potential's minimum because of the huge strain that would otherwise be imposed. 

To find clusters\footnote{Hereafter we say ``cluster'' to mean ``rigid cluster'' unless context makes it clear otherwise.} with the maximum possible number of contacts $M(n)$ for $n=$ 15--19 we make the following approximations. First, we restrict computations to the set  of clusters with at least  $m$ contacts; call this set $S_m$. Second, we break no more than 3 contacts simultaneously  to find each transition path. 
These truncations are reasonable because we checked that for $n\leq 13$, $S_m$ is  connected provided $m > M(n)$, where $M(n)$ is the maximum possible number of contacts, and this continues to hold even when breaking no more than 3 contacts.

Table \ref{tbl:results} organizes the data according to the number of contacts in each cluster. A generic rigid cluster must have $3n-6$ contacts, and indeed most clusters do: 97\% for $n=11-14$. However, our system is not generic because the spheres all have the same diameters, so there are many solutions that are special in some way. 
We now highlight these and other interesting features of the dataset.

\begin{figure*}
\begin{center}
\includegraphics[width=1\linewidth]{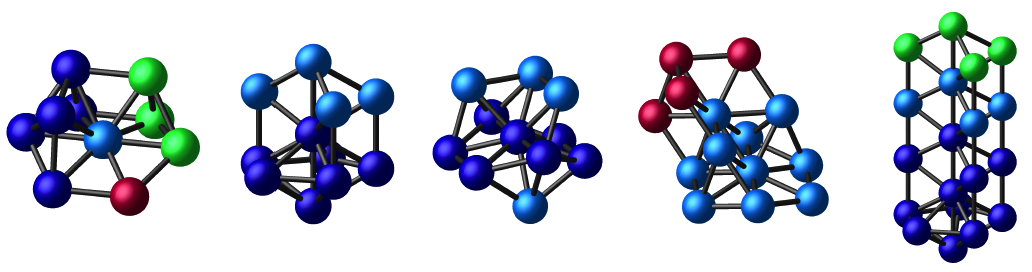}
\caption{Selected hypostatic clusters. From left to right: $n=10$ (smallest hypostatic cluster, missing 1 contact), $n=11$ (missing 2 contacts), $n=11$ (missing 2 contacts), $n=14$ (missing 3 contacts; light blue spheres are $n=10$ hypostatic cluster), $n=19$ (missing 6 contacts; dark blue spheres are $n=11$ hypostatic cluster).}\label{fig:hypostatic}
\end{center}
\end{figure*}

\paragraph{ \bf Hypostatic clusters}

The biggest surprise is the existence of clusters with \emph{fewer} than $3n-6$ contacts: hypostatic clusters. The smallest has $n=10$ spheres and is ``missing'' one contact; it is shown in Figure \ref{fig:hypostatic} (left). To our knowledge this is the first discovery of a hypostatic packing of spheres. (Hypostatic packings of ellipsoids have been observed \cite{donev2004},
but these arise because of the extra rotational degrees of freedom when the aspect ratio of the sphere is perturbed \cite{donev2007}.) For $n=11$ there are 2 clusters missing two contacts, also shown in Figure \ref{fig:hypostatic}, and 18 clusters missing one contact. Most clusters missing one contact are obtained from the $n=10$ hypostatic cluster by gluing a sphere, but the clusters missing two contacts cannot be formed this way. 
The smallest cluster missing 3 contacts occurs for $n=14$, see Figure \ref{fig:hypostatic}. 

All of these are rigid, despite having several linear degrees of freedom.
It is helpful to build these using a magnetic ball-and-stick set,\footnote{For example as manufactured by Geomags or CMS Magnetics.} because the rigidity is a highly cooperative property that is hard to convey in two dimensions. 
A common feature is that several spheres lie in a plane stabilized at its boundary. Spheres in the plane can move perpendicular to it infinitesimally, but cannot move any finite amount without breaking a contact somewhere else. 
For example, the $n=10$ cluster is made of a rigid 6-cluster (dark+light blue) and a tetrahedron (green+light blue) that share a sphere (light blue) and two contacts to make a square face. Without the red sphere, these can twist along a single degree of freedom. When the red sphere is added exactly in the plane of the spheres it contacts, this stabilizes the twist and rigidifies the cluster. 

\begin{figure}
\begin{center}
\includegraphics[width=0.6\linewidth]{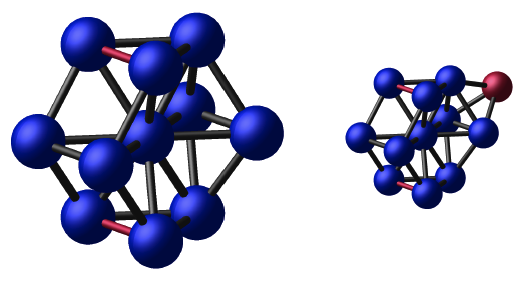}
\end{center}
\caption{A cluster the algorithm cannot find (left) (n=11). This is a fragment of an hcp lattice that is missing 1 contact. Breaking either or both of the red contacts gives a nonlinearly rigid cluster. Breaking any other contact gives a cluster with two degrees of freedom. Therefore there are no one-dimensional transition paths leading to this cluster. 
Right: the cluster it came from (n=12), by deleting the red sphere. 
}\label{fig:lost}
\end{figure}

Not all hypostatic clusters are built on planar arrangements though; see the cluster in Figure \ref{fig:lost} which will be discussed later.

One might wonder: is it possible to build a rigid cluster missing arbitrarily many contacts? The answer is yes, and can be done using the left-most $n=11$ cluster as a scaffold. This cluster is formed from a rigid cluster of 7 spheres (dark blue) combined with a group of 4 (light blue) arranged in three planes that intersect in a line. One can continually add groups of 4 on top of this to extend the planes; each group of 4 increases the number of contacts missing by 2, so that asymptotically the number of contacts is $\sim 2n$ as $n\to \infty$. Figure \ref{fig:hypostatic} shows a cluster of $n=19$ that is missing 6 contacts. 

In fact, Robert Kusner \cite{kusner} pointed out a family of rigid $n$-clusters where the number of contacts grows asymptotically as $\sim n$: Construct a rigid $n$-cluster from a large, ball-like region of the face-centered cubic (fcc) sphere packing, carving out the inside in such a way that the remaining spheres form a rigid shell surrounding a large cavern crossed by many parallel columns of spheres from the fcc packing.  Since the number of spheres in this cluster is dominated by the number of spheres in the columns, which scales like the volume of the cavern, while the remaining number of spheres scales like the area of the shell, the number of contacts is asymptotically that for the columns, namely, $1$ per sphere.  This is the smallest possible asymptotic growth rate for the number of contacts, since otherwise the cluster loses connectivity.

\begin{figure*}
\begin{center}
\includegraphics[width=1\linewidth]{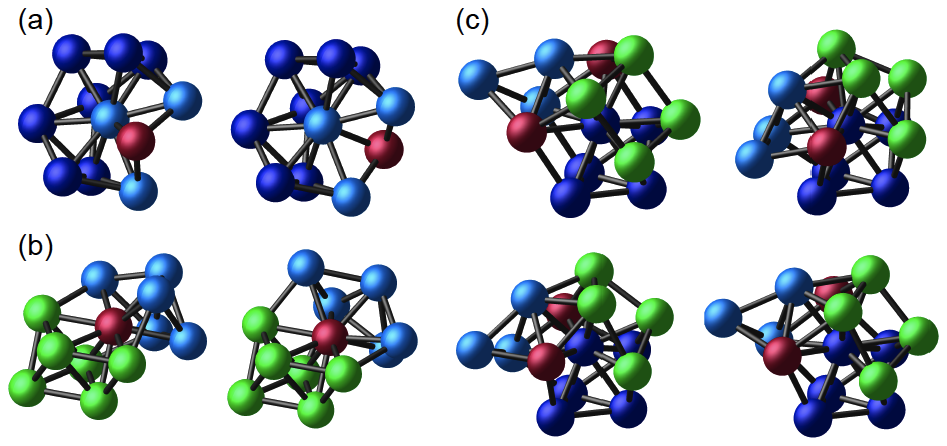}
\end{center}
\caption{(a) Smallest clusters with the same adjacency matrix  ($n=11$) . The dark(light) blue spheres have identical coordinates on each, while the red sphere forms three contacts with the light blue spheres in two different ways. 
(b) Another pair with the same adjacency matrix  ($n=12$). The green+red spheres form a rigid cluster and have identical locations in each cluster, while the blue+red forms another rigid cluster. It shares three contacts and the red sphere with the green+red cluster, and there are two different ways to perform this gluing. 
(c) Four clusters with the same adjacency matrix ($n=14$) . The five dark blue spheres form a rigid cluster, while other spheres are colored to aid visualization. }\label{fig:adj}
\end{figure*}

\paragraph{ \bf Clusters with the same adjacency matrices}

Another big surprise is that sometimes  \eqref{eq:bonds},\eqref{eq:cons}  has more than one physically realizable solution. In this case a single adjacency matrix corresponds to two or more distinct clusters. The smallest cluster for which this occurs is $n=11$ and the pair of solutions is shown in Figure \ref{fig:adj}. 
They differ by the location of a single sphere  (red), which forms contacts with three spheres (light blue) either above or below the plane of the these spheres. 

For $n=12$ and $13$ we find 23 and 474 pairs with the same adjacency matrix respectively. For $n=14$ there are 666,3,3 adjacency matrices with 2,3,4 solutions respectively. 
These multiple solutions do not all differ by a single sphere but can vary in more complicated ways. For example, Figure \ref{fig:adj} 
shows a pair of solutions for which all spheres have at least four contacts. This solution is made of two rigid clusters of 6 (light blue / red) and 7 (green/red) spheres, glued together so they share a sphere (red) and form 3 other contacts. One can check this gluing condition implies the resulting cluster has $3n-6$ contacts if the two sub-clusters do. The figure also shows a quadruple of solutions for $n=14$, which is hard to decompose into rigid body sub-components. 

Note that generically, for an infinitesimally rigid cluster with no inequality constraints and random edge lengths, one actually expects to have a very large number of different embeddings -- the best-known upper bound on the number of embeddings grows exponentially with $n$, and one can construct classes of two-dimensional graphs where the lower bound does also \cite{streinu2004}. Therefore, the surprise could equally be construed as there being so \emph{few} cases where \eqref{eq:bonds},\eqref{eq:cons} have multiple solutions. This points to the importance of non-overlap constraints, and the related phenomenon of geometrical frustration, in physical systems.

\paragraph{\bf Hyperstatic clusters / Ground states} 

\begin{figure*}
\begin{center}
\includegraphics[width=1.08\linewidth]{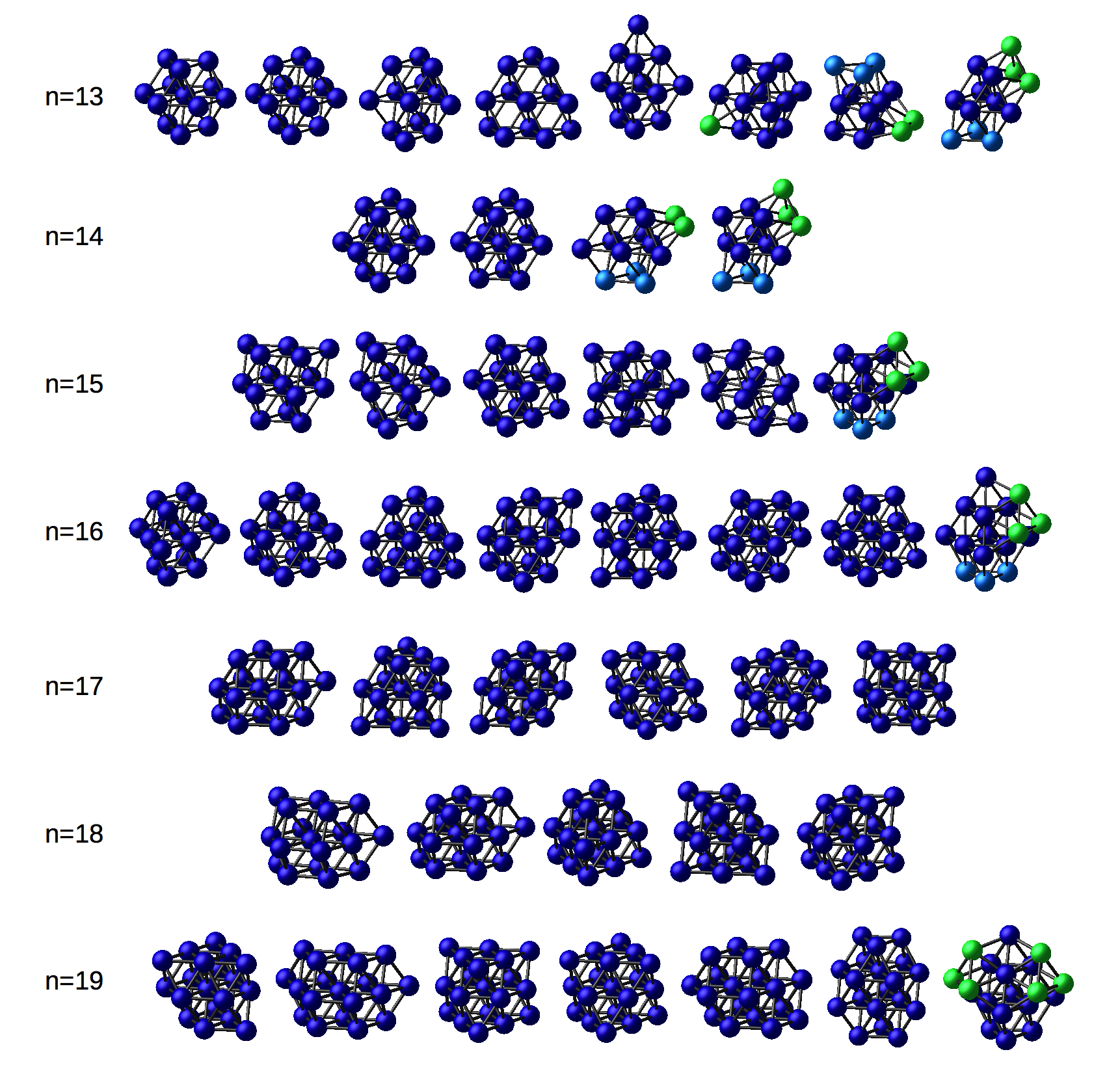}
\end{center}
\caption{ Clusters with the maximum number of contacts, for $13\leq n\leq 19$. Dark blue spheres all form a close-packing, while coloured spheres form defects. In some cases there is more than one natural way to group the spheres so the defects are given different colours, i.e. the close-packing is either (dark blue + light blue), or (dark blue + green). 
}\label{fig:max}
\end{figure*}

At $n=10$ we begin to find clusters with \emph{greater} than $3n-6$ contacts: hyperstatic clusters.
This is expected, since the densest known sphere packing of $\mathbb{R}^3$ has spheres with 6 contacts each, arranged in a close-packing by stacking planes of a hexagonal lattice \cite{hales2005}. 

The ground-state clusters, those with the maximum possible number $M(n)$ of contacts, are shown for selected $n$ in Figure \ref{fig:max}. 
Dark blue spheres form 
fragments of a close-packing,  and other colours represent defects. 
The ground states appear to converge rapidly to close-packings: for $n=15,16,19$ there is one ground state with a defect, and for $n=17,18$, there are none. This is in marked contrast to clusters with a smoother potential such as Lennard-Jones clusters, where non-close-packed arrangements are common in low-energy clusters for $n$ up to at least $100$ \cite{walesdoye1997}.  

A famous mathematical problem is the ``kissing problem,'' which asks how many spheres can touch a single sphere without overlap. The answer is 12 \cite{conway1999}, and the two clusters that achieve this, one a piece of a hexagonal close-packed (hcp) lattice, the other of a face-centered cubic (fcc) lattice, are the left-most ground states for $n=13$.  What is not commonly known is there are six other clusters with the same number of contacts, three of which have defects. As $n$ increases, ground states more frequently contain caged spheres: 2/4, 5/6, 6/8, 6/6, 5/5, 6/7 for $n=$14--19 respectively.

\paragraph{\bf Close-packing fragments} Overall, the percentage of clusters which are fragments of a close-packing decreases with $n$, as $17\%,7.2\%,3.6\%,1.6\%, 0.63\%$ for $n=$10-14 respectively. 
A greater proportion of hyperstatic clusters are close-packing fragments \cite{SI}. 
Some of these are fragments of an hcp or fcc lattice;  
the ground states have slightly more hcp than fcc fragments, though we expect random stackings (neither hcp nor fcc) to dominate as $n$ increases as the stacking pattern should be combinatorially favoured. 
It is interesting that spheres in condensed-matter systems frequently crystallize to form a close-packed lattice, despite an apparently increasingly large entropy favouring other states, particularly non-ground-states. This dataset and the associated rearrangement and growth pathways could be used to understand such nucleation phenomena. 

\paragraph{\bf Singular clusters}

\begin{figure}
\begin{center}
\includegraphics[width=0.8\linewidth]{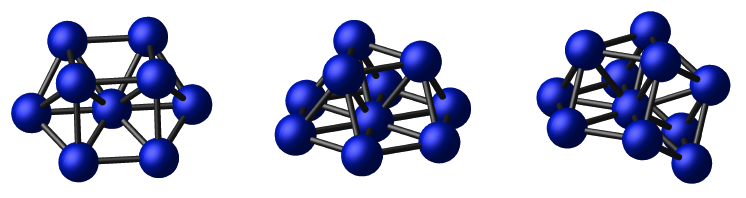}
\end{center}
\caption{ Selected singular clusters, for $n=9,10,11$ (left to right). 
Each has a linear degree of freedom, but is nonlinearly rigid.
These are new seeds: they cannot be formed by gluing spheres to a smaller cluster. They all have $3n-6$ contacts. }\label{fig:singular}
\end{figure}

The dataset contains a great many singular
clusters: those which are linearly floppy, but nonlinearly rigid\footnote{Usually ``singular'' also includes all clusters with extra contacts, but here we use the word to mean a more restrictive set.}.
All hypostatic clusters are singular, but many others are as well including those with $\geq 3n-6$ contacts.
The smallest occurs at $n=9$ and is shown in Figure \ref{fig:singular} 
(left); this is a fragment of an fcc lattice. As $n$ increases the fraction of singular clusters is nearly constant:  $3\%,2.9\%,2.7\%,2.5\%$ for $n=11,12,13,14$ \cite{SI}. 
Hyperstatic clusters can also be singular, as are two of the ground states for $n=13$.

The $n=9$ singular cluster was observed to occur experimentally with surprisingly high probability. These experiments considered  isolated systems of $N=9$ (or smaller) micron-scale hard spheres in a fluid, interacting with a short-range potential induced by depletion forces. They measured the equilibrium probabilities of all 52 clusters \cite{meng2010}, and found the singular one with frequency 11\%. This is much higher than if all clusters occurred with equal probability, and is not explained by considering particle permutations, or rotational or translational entropies. What is a likely candidate is the vibrational entropy --- the volume of configuration space explored by fast particle vibrations induced by the short-range potential. For a regular cluster, one can approximate this entropy by supposing the spheres in contact are attached by harmonic springs, so the energy near the cluster is quadratic. For a singular cluster, this approach fails, because one of the directions of the quadratic function becomes perfectly flat so the entropy associated with this direction diverges. 

Therefore, singular clusters may be an important part of the free-energy landscape when the spheres are thermal. Since they appear to be robust (at least for small $n$), it is important to learn how to evaluate the entropies of these clusters correctly. 
Ideally one would like the result to be independent of the choice of interaction potential, for example as in \cite{holmescerfon2013}. 
While it is not yet known how to do this beyond the harmonic approximation, it is possible that ideas from algebraic geometry involving asymptotic volumes of intersecting manifolds may be helpful  \cite{tschinkel2010}, provided they can also be turned into a computational algorithm that can be applied to systems with many variables.

\begin{figure}
\begin{center}
\includegraphics[width=1\linewidth]{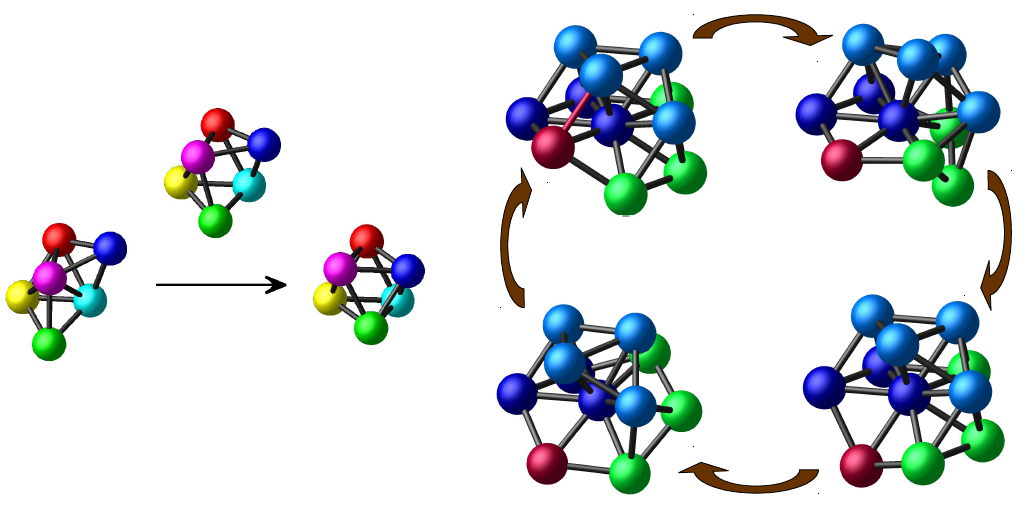}
\end{center}
\caption{ Left: a typical transition path, with the topology of a line segment. A contact is broken, and the cluster deforms into a geometrically distinct cluster. 
Right: a circular transition path ($n=11$.) When the red contact breaks (top left), the light blue spheres twist to the right (top right), creating space for the red sphere to move down past the plane of the dark blue spheres, which are fixed in place. When the light blue spheres twist back to center all spheres return to their original positions except the red one (bottom right). The light blue spheres twist left (bottom left), allowing the red sphere to move back up through the plane and return to its original position. 
}\label{fig:circle}
\end{figure}

\paragraph{\bf Circular transition paths}
The paths used to find rigid clusters are also relevant pieces of a cluster's configuration space. For example, they are the minimum-energy paths between rigid clusters when the spheres interact with a short-range potential, so can be used to evaluate the leading-order transition rates \cite{holmescerfon2013}.
What are the topologies of the paths? Most are intervals connecting distinct clusters, as expected (Figure \ref{fig:circle} 
(left).) Interestingly, we also find paths which are circles: we break a contact, move along the one-dimensional path, and re-form exactly the same contact in exactly the same configuration. This first happens for $n=11$ as shown in Figure \ref{fig:circle} 
(right). For $n=13$ there are roughly $18,300$ circular paths, and they occur for all types of clusters and paths (regular/singular/hyperstatic/hypostatic.) 

This suggests there could be circular paths that  \emph{do not} eventually form another contact: the cluster may internally deform indefinitely without ever becoming rigid. 
Such a cluster would be metastable since it could not reach a lower-energy state without first breaking a contact, so for certain applications one should treat them on par with rigid clusters.  How to find such circular metastable states is an open question. 

We note in passing that these paths may be analogous to the ``localized'' soft modes recently discovered in gels and networks  \cite{lubensky2014,paulose2014}, where an infinite collection of points with distance constraints contains an apparently circular deformation mode with nearly localized point displacements.

\paragraph{\bf A cluster with no one-dimensional deformation paths} 
We found a cluster at $n=11$ that cannot be accessed from any other by one-dimensional paths, shown in Figure \ref{fig:lost}. 
We checked that breaking any single contact except two (in red) forms a regular cluster with two degrees of freedom. 
When either or both of the red contacts is broken, the cluster is still rigid.
Therefore there does not exist a subset of contacts that, when broken, forms a cluster with one degree of freedom.  

We found this cluster by accident, by deleting a sphere (red) from an $n=12$ cluster. 
This leads to an intriguing idea --- perhaps other such highly singular clusters could be discovered through this method of ``catalysis'' by extra particles?


\section{Discussion}

\subsection{How to understand hypostaticity?} 
\begin{figure}
\begin{center}
\includegraphics[width=1\linewidth]{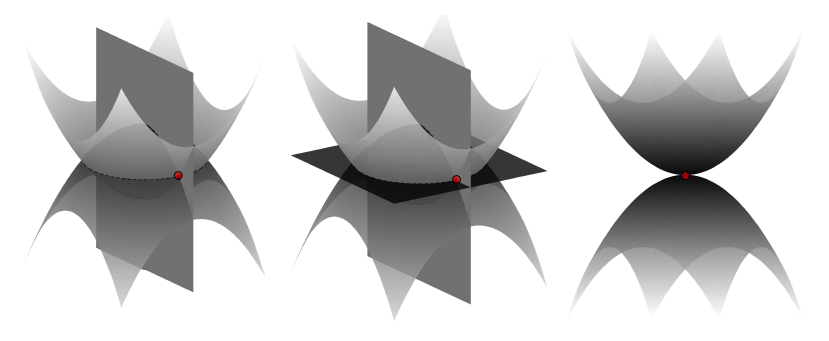}
\end{center}
\caption{A low-dimensional analogy to understand nonlinear rigidity. Each contact is represented by a surface, and a nonlinearly rigid cluster is an isolated intersection point (in red).
Left: generically, three surfaces are required in $\R^3$ for an isolated intersection point; this is an analogy for a regular cluster. 
Middle: four surfaces intersecting at a single point, as for a hyperstatic cluster. 
Right: two surfaces that intersect at a single point, as for a hypostatic cluster. 
 }\label{fig:schem}
\end{figure}

How is it possible for a cluster to have fewer than $3n-6$ contacts?  
To answer, it is helpful to have a low-dimensional analogy of system \eqref{eq:bonds}. 
Suppose the space of ``clusters'' is three-dimensional. Then each single contact equation in \eqref{eq:bonds} is solved by points on a two-dimensional manifold: a surface. Two contacts make two surfaces that generically intersect in a one-dimensional manifold, or a curve, and three contacts make three surfaces that generically intersect at a point, as shown in  Figure \ref{fig:schem}
(left).

Following this analogy, having extra contacts means there is one (or more) additional equation in \eqref{eq:bonds} for which $ \bb{x}$ is already a solution.  Figure \ref{fig:schem}
(middle) shows an example where four surfaces intersect at a point instead of three. Deleting one of the surfaces still leads to an isolated intersection point.

Having fewer contacts is also possible, as shown in Figure \ref{fig:schem}
(right). Here two surfaces intersect at a single point where they are exactly tangent. Linear analysis would predict a two-dimensional solution set, however we cannot move any small but finite distance away and remain on both surfaces, so this point is a rigid solution. Perturbing the surfaces slightly, for example by shifting one up or down, will either destroy the intersection point or turn it into a one-dimensional curve, so a hypostatic cluster (and any singular cluster) is a special case that is extremely sensitive to the choice of parameters. 

\subsection{Comparison with previous results}

Two previous studies sought to enumerate clusters with $3n-6$ contacts, by listing all non-isomorphic adjacency matrices and trying to solve \eqref{eq:bonds} for each. Arkus et al. \cite{arkus2011} solved for the positions of the sphere centers analytically, or proved there is no solution, so this is guaranteed to be the complete set of clusters up to $n=10$ with at least $3n-6$ contacts, although these were not tested for rigidity. 
Hoy et. al. \cite{hoy2012} used Newton's method with random initial conditions to solve for the positions of the sphere centers. 
Their method is not provably complete, partly because Newton's method is not guaranteed to find a solution even if it exists, and partly because they incorrectly assumed a rigid cluster always contains a Hamiltonian path \cite{hayes2013}. 
Nevertheless they produced a large dataset for $n\leq 11$ that can be used to test other methods.\footnote{Since this paper was submitted, one more enumeration study has been published \cite{hoy2015}. This uses the same method as \cite{hoy2012} without the Hamiltonian path assumption, to list clusters of $n\leq 13$.} 
 Our set of clusters is nearly identical to theirs, but with the following discrepancies for $n=11$: 
(i) They list two clusters as rigid that our method identifies as floppy. 
(ii) They do not find the second solution for the adjacency matrix with two solutions. 
(iii) They do not find hypostatic clusters, because they do not look for these.

\subsection{Completeness of the data}
The discovery of a cluster that has no one-dimensional deformation paths implies the algorithm cannot find all rigid clusters --- there may be other single clusters or collections of clusters that are not accessible from the starting set by such paths. 
Additionally, we test only for pre-stress stability, and do not consider clusters that are rigid under weaker conditions. 
Nevertheless, the agreement with previous studies for smaller $n$ makes it reasonable to expect it finds the vast majority of rigid sphere packings; we suspect the missing clusters or collections are rare. 
Perhaps it enumerates a complete subset of packings: for example,  the set of non-singular clusters may be connected by one-dimensional transition paths. Proving such a statement or a variant (such as for non-identical spheres, or overlapping spheres), would be valuable because the algorithm is fast, so it could be applied to less-studied systems such as spheres of different sizes, or objects of different shapes.

\subsection{Bottleneck} 
The computational bottleneck is not only the factorial growth in the \emph{number} of clusters, but also the growth in the \emph{number of subsets of contacts} that must be broken to find transition paths leading out of hyperstatic clusters. 
For example, for a cluster of $n=13$ with 3 extra contacts, we have to break 4 contacts generically to find a transition path, so we have to check roughly ${3n-3 \choose 4}$ sets. This is the equivalent of $\approx 2,000$ regular clusters, or $1/50$ of the total number. This worsens at the number of extra bonds grows: for $n=19$, a ground state with 9 extra contacts is the equivalent of roughly ${3n+3\choose 9}/(3n-6)\approx 10^9$ regular clusters.
Higher $n$ may be feasible by developing new methods to avoid checking all possible subsets, e.g. by predicting in advance which subsets are likely to lead to valid transition paths. Ideas from physics may be useful, for example those which study the ``soft'' modes of interacting particle systems and show these give rise to catastrophic deformations \cite{wyart2005,wyart2012}. 

\subsection{Weaker rigidity tests} 

It would be fascinating to find clusters that are rigid under a weaker condition than pre-stress stability, but there is no known efficient method to test this. 
Indeed, even the notion of higher-order rigidity is difficult to define \cite{connelly1994, garcea2005}.
Yet, to compute clusters for physical applications, it is not always necessary to prove rigidity, numerically or otherwise. 
Sometimes, knowing that a cluster is rigid or floppy ``to a certain order'' is useful information, because it may be that higher-order physics controls the flexibility properties beyond this. 
It would be interesting to extend rigidity tests to include physically relevant higher-order information, despite what these do or do not imply for actual rigidity. 


A related problem is to consider tests for floppy clusters. Is there an equivalent notion of pre-stress stability for a cluster with one or more internal degrees of freedom -- i.e. is there a test that guarantees the cluster lives on a manifold of a particular dimension? Answering this would not only make the algorithm above more rigorous, but would also aid in computing more general configuration spaces, which can include floppy components \cite{sitharam2011}. 

Questions such as these arise in more general systems than clusters. For example, \cite{paulose2014} has found a network of linkages (points connected by distance constraints), with a mode of deformation that appears only in the limit as the system size approaches infinity. 
The mode is localized, with displacements decreasing exponentially away from a hot spot, yet these very small displacements are enough to cause the system to be technically rigid for any finite system size. 
Whether the mode is a finite or an infinitesimal motion determines how the network behaves as a material, but it is not yet understood how to determine this within the current rigidity framework. 

It is worth noting that  \eqref{eq:bonds},\eqref{eq:cons} is a system of algebraic equations, so in principle we can extract any information about the solution, such as whether it is an isolated root, a multiple root, or a positive-dimensional solution, by knowing its full algebraic properties. We have tried implementing known algorithms --- such as computing the Gr\"{o}bner basis \cite{grobner}, Hilbert function \cite{dayton2005,dayton2011,wampler2011}, or solving directly using the numerical algebraic geometry package Bertini \cite{wampler2013}, but the system was too large for all of these. For certain clusters these worked when we could reduce the system size, for example by decomposing the cluster into rigid sub-clusters that are already known, but we have not yet found a way to make this work for the full generality of clusters encountered. 

\section{Conclusion}

We computed a set of  rigid clusters of $n\leq 14$ hard spheres, using a dynamical algorithm that follows all possible transition paths between clusters. There is empirical evidence to believe this is the entire set of such clusters, minus a few rare clusters that are either too singular or cannot be reached by one-dimensional transition paths. This dataset is much larger and more complete than those produced before, so we expect it to be useful in addressing questions in physics, chemistry, and materials science, for example. Along the way we raised several issues that could usefully be addressed from a mathematical, geometrical, or computational perspective. 
The data contains several surprises, perhaps the biggest of which is 
clusters that are hypostatic. The thermodynamic and material importance of these is currently unknown. 
We discussed these and other clusters, which are hoped will stimulate the mathematical and material imaginations. 

\bigskip

\begin{acknowledgments}
Thanks to Michael Brenner, Steven Gortler, and Vinothan Manoharan for many helpful discussions, and to Robert Hoy for sharing his data. 
This material is based upon work supported by the U.S. Department of Energy, Office of Science, Office of Advanced Scientific Computing Research
under award DE-SC0012296.
\end{acknowledgments}


\bibliography{/Users/miranda/Dropbox/Work/Bibliographies/ColloidBib.bib}

\clearpage


\appendix*

\section{Appendix}


\section{Data}

This section contains more detailed statistical information about the set of clusters. 
The set of coordinates is listed on the author's website \cite{mhcwebsite}.

\subsection{Total number of clusters}

\begin{figure}
\begin{center}
\includegraphics[width=1\linewidth]{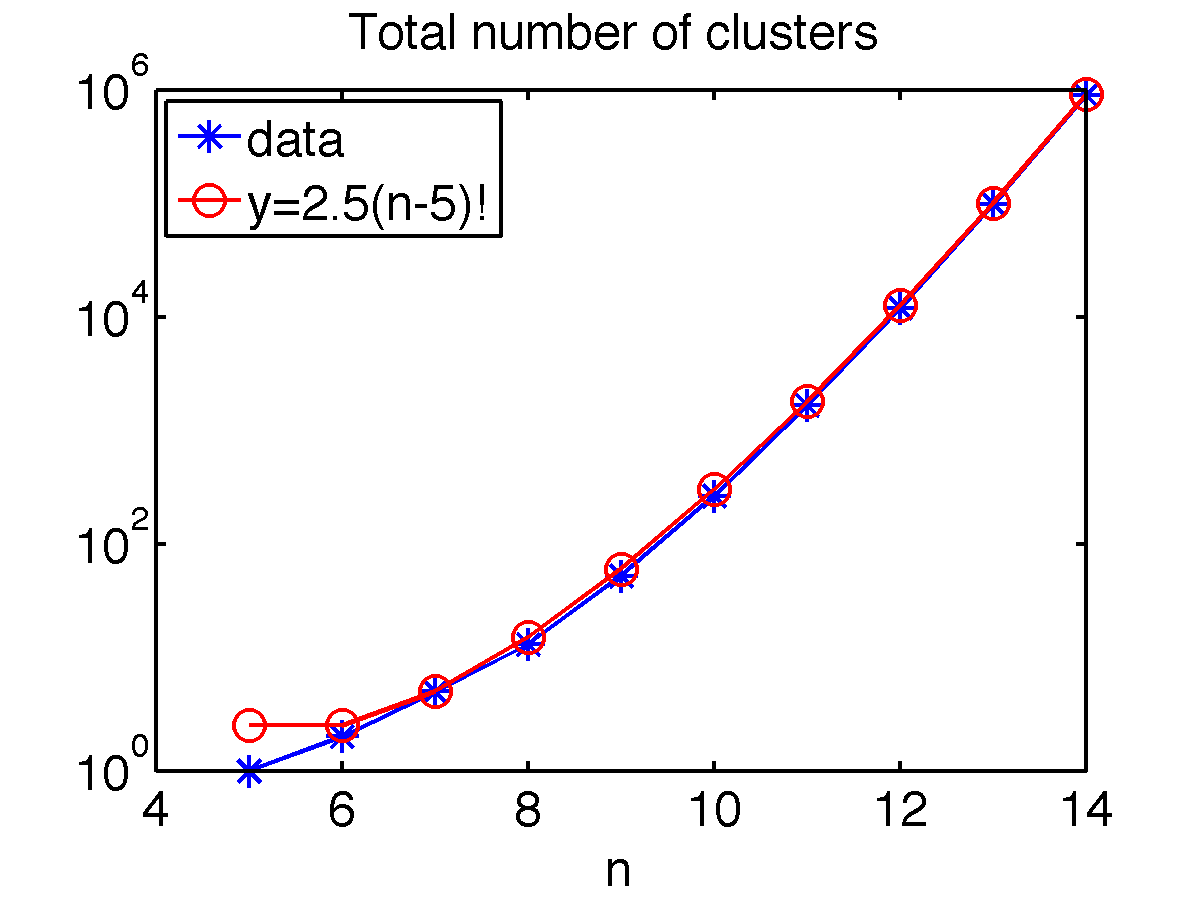}
\end{center}
\caption{Total number of clusters $N(n)$, and the best-fit curve $2.5(n-5)!$}\label{fig:Nn}
\end{figure}

Consider the ratios of total number of clusters $N(n+1)/N(n)$. Table \ref{tbl:ratios} shows this appears to be multiplied by nearly $n-4.8$ for each $n$. 
Therefore the total number grows roughly as $(n-5)!$. We fit the total number $y = b\cdot(n-5)!$. Minimizing the mean-square error in linear space over $5\leq n \leq 14$ gives $b=2.46$. This does a very good job for the data computed, (see Figure \ref{fig:Nn}), but will likely slightly underestimate for larger $n$. 

\begin{table*}
\begin{tabular}{c | ccccccccccccc | c}
$n$ & \multicolumn{13}{c|}{\# of contacts - $3n$} \\ 
  & $-9$ & $-8$ & $-7$ & $-6$ & $-5$ & $-4$ & $-3$ & $-2$ & $-1$
   & $0$ & $+1$ & +2 & +3 &Total \\ \hline
6 && & & 2 &&&&&&&&&& 2\\
7 && & & 2.5  &&&&&&&&&& 2.5\\
8 && & & 2.6  &&&&&&&&&& 2.6\\
9 && & & 4  &&&&&&&&&& 4\\
10 && & -- & 5.0 & --  &&&&&&&&& 5.1 \\  
11 && -- & 18 & 6.2 & 6.7 & --  & & & &&&&& 6.3 \\
12 && 5.5 & 8.2 & 7.2 & 8.7 & 8 & -- & &&&&&&  7.2 \\
13 && 7.9 & 8.3 & 8.2 & 7.5 & 12 & 8 & & &&&&& 8.2 \\
14 &--& 7.3 & 8.6 & 9.1 &  7.9 &  9.1 &  9.9 & -- & &&&&& 9.1 \\\hline
15 &&    &    &    &   & 8.7  & 9.9  &  13.8 & -- & &&& &\\
16 &&    &    &    &   &   &  10.1 & 12.1 & 10.3 & -- &&& &\\
17 &&    &    &    &   &   &   & 11.7 & 12.7 & 10.6 &-- && & \\
18 &&    &    &    &   &   &   &  & 12.1 & 12.8 & 15.2 & -- && \\
19 &&    &     &   &   &   &   &  &  & 12.4 & 16.0 & 19.0 &-- & \\ \hline
\end{tabular}
\caption{Ratios of number of cluster of each type, to the number of the same type with one less sphere. }\label{tbl:ratios}
\end{table*}

\subsection{Close-packing fragments}

We determined whether a cluster was a close-packing fragment by choosing three mutually contacting spheres, and rotating this triangle to the seven triangles of a bipyrapid (two tetrahedra in contact.) For each rotation we checked whether all the $z$-coordinates of the cluster were integer multiples of $\sqrt{2/3}$, and if they were we checked whether each plane of spheres at constant $z$ formed a hexagonal lattice. 
The data is shown in Table \ref{tbl:lattice}. This also identifies the lattice type of clusters in Figure \ref{fig:max} in the main text. 

\subsection{Singular clusters}

We computed the number of rigid clusters for which the Jacobian of \eqref{eq:bonds} had at least one element in the null space. This data is shown in Table \ref{tbl:singular}. We also show the numbers of clusters which are both singular, and close-packing fragments. 


\subsection{Gaps}

As $n$ increases, the smallest gap between spheres not in contact decreases. 
This is one major difference between sphere packings and clusters with a smooth potential, and may be part of the explanation for the combinatorial rather than exponential growth in the number of clusters. 

We checked that we are resolving the smallest gap by computing the minimum gap size over all clusters, where the gap for each cluster is the minimum pairwise distance between non-contacting spheres minus 1. (Recall that for $n\geq 15$ not all clusters are computed.)  The minimum gap decreases continually with $n$ as follows:\\

\hspace*{24pt}\makebox[\linewidth][l]{
\begin{tabular}{cl}
$n$ & minimum gap \\\hline
6 & 0.4142\\
7 & 0.05146  \\
8 & 0.05146  \\
9 &  0.05146 \\
10 & 0.03296  \\
11 & 0.02634  \\
12 & 2.129e-3  \\
13 & 5.768e-5  \\
14 & 1.269e-5  \\\hline
15 &  0.004364 \\
16 & 0.006154  \\
17 & 0.006154  \\
18 & 0.006154 \\
19 & 0.006154 \\\hline
\end{tabular}\\
}

\bigskip
 
At $n=13$ there are 23, 9 clusters with gaps less than $10^{-3}, 10^{-4}$ respectively. These are all regular clusters with $3n-6$ contacts. The ten smallest gaps are $10^{-5}\times$ (5.768,  6.881,  7.339,  7.339,  7.361,  7.505,  7.507,  7.635, 7.694, 11.02). 

At $n=14$ there are 929, 244, 34, 6 clusters with gaps less than $10^{-3}, 10^{-4}, 5\times10^{-5}, 2\times10^{-5}$ respectively. 
All have $3n-6$ contacts, and all are regular except three of the ones with the largest gaps. The ten smallest gaps are $10^{-5}\times (1.269, 1.377, 1.385, 1.385, 1.387, 1.744, 2.536, 2.538, 2.538, 2.539)$.  

The smallest gaps for $n=14$ are close to our tolerance for adjacency (\texttt{tolA} = $10^{-5}$), but the gaps appear to approach the minimum smoothly, with a jump to \texttt{tolA},  so it is likely we have resolved the cutoff. However, applying the algorithm for larger $n$ will require changing the numerical parameters to resolve smaller gaps. 

\begin{figure}
\begin{center}
\includegraphics[width=0.48\linewidth]{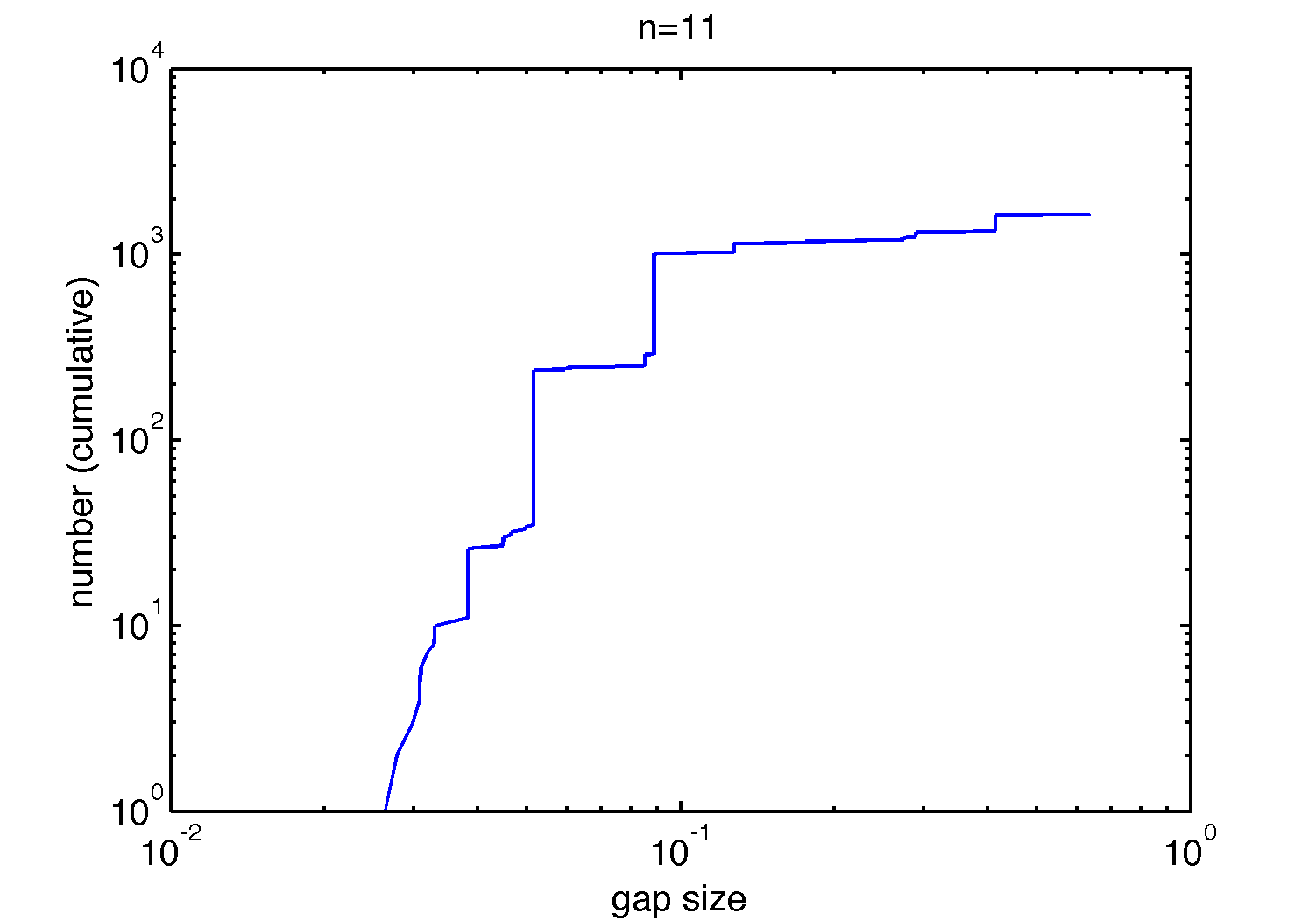}
\includegraphics[width=0.48\linewidth]{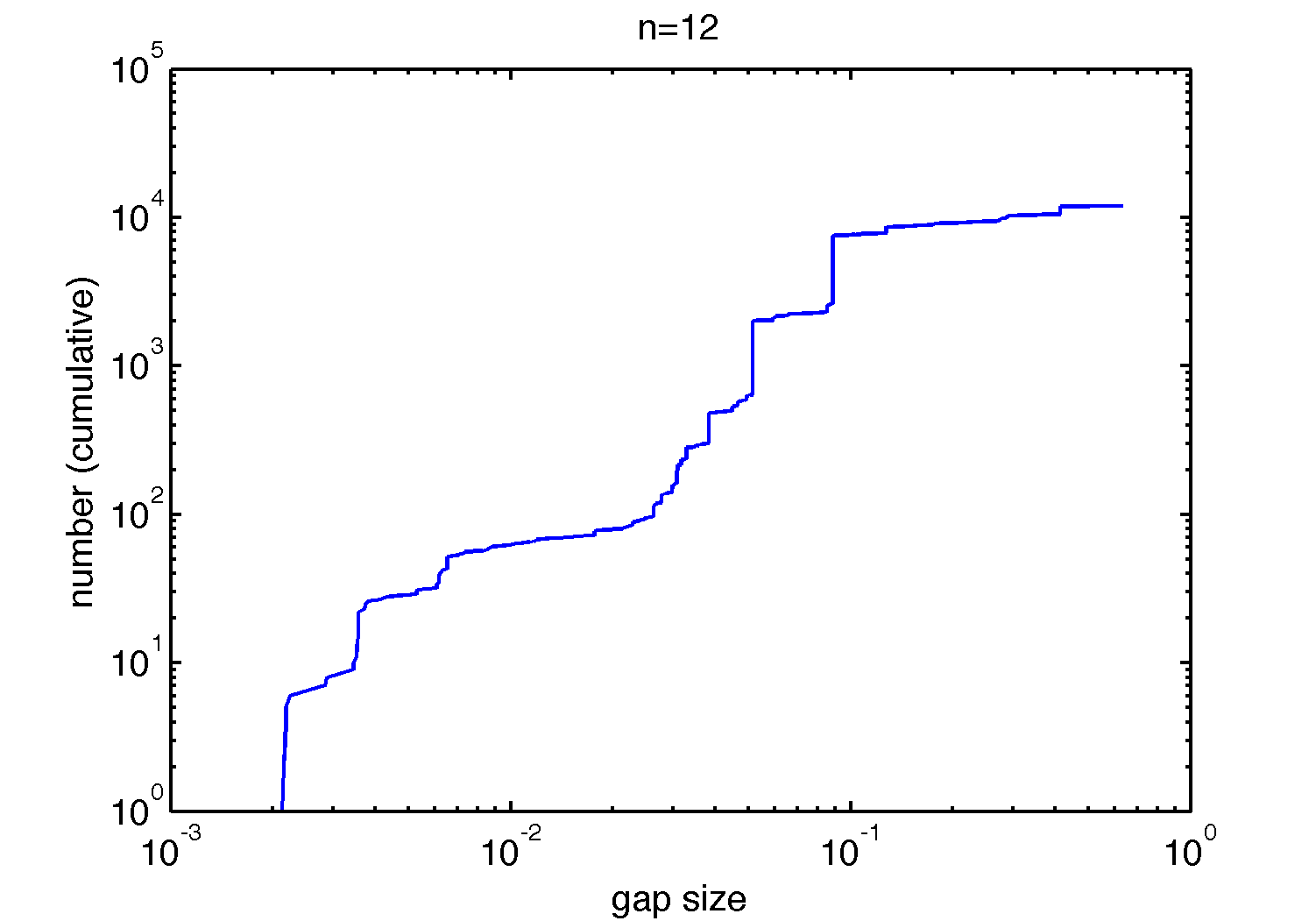}\\
\includegraphics[width=0.48\linewidth]{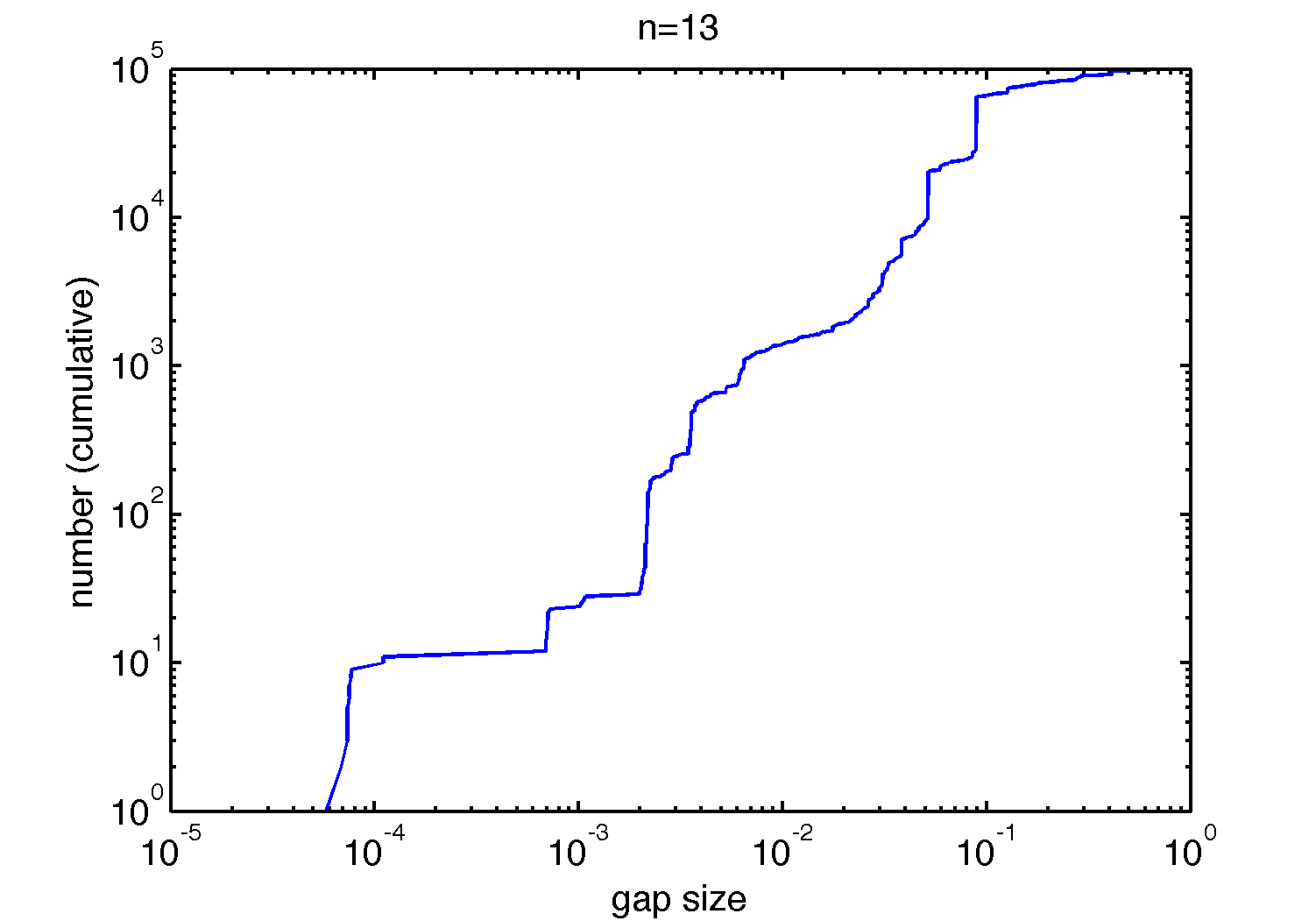}
\includegraphics[width=0.48\linewidth]{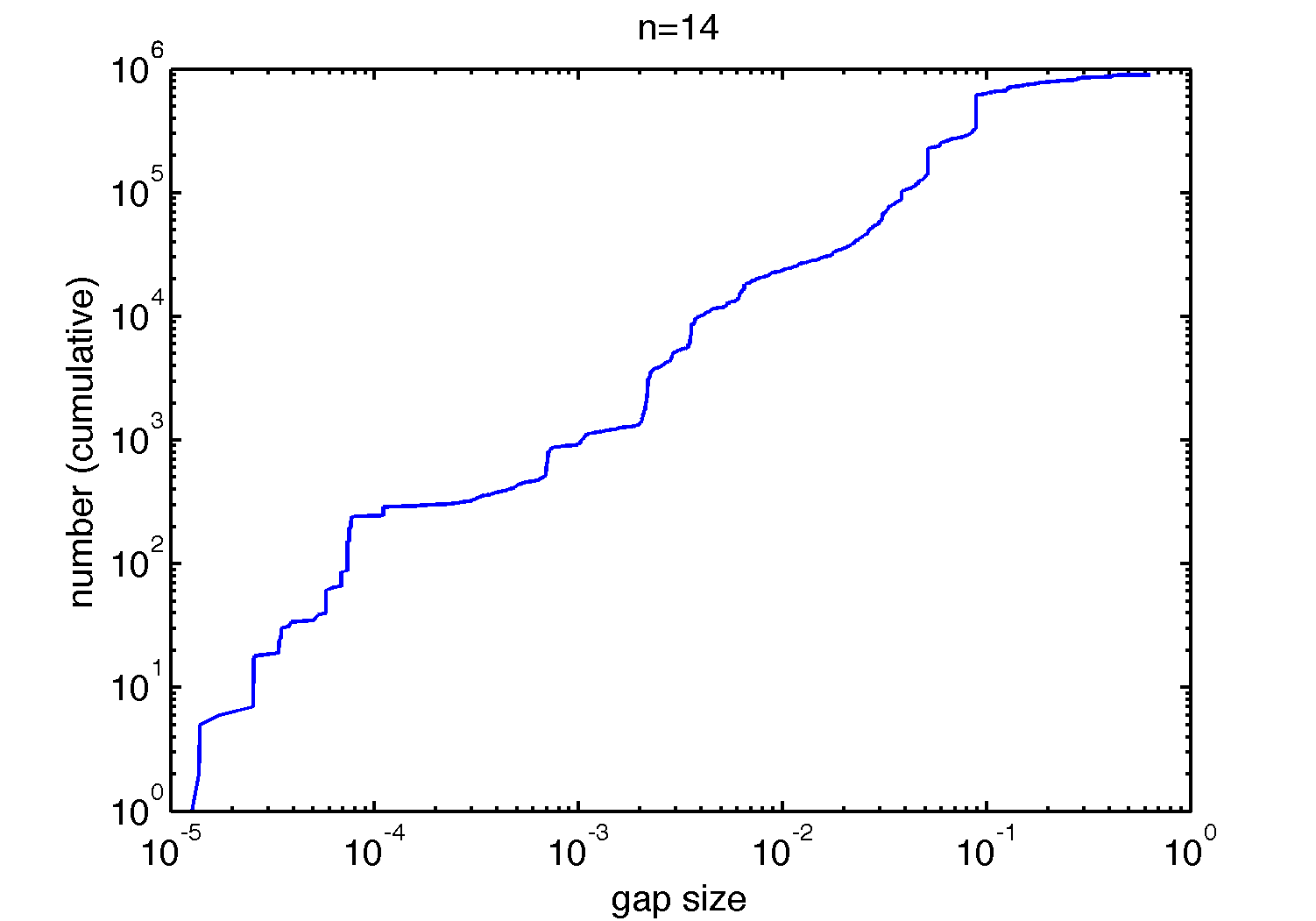}
\end{center}
\caption{Cumulative gap size distribution for $n=11,12,13,14$. }\label{fig:gaps}
\end{figure}

The cumulative gap size distributions are shown in Figure \ref{fig:gaps}. These are a fascinating mixture of smooth distributions, plus sharp jumps when many clusters have the same minimum gap size.

\subsection{Scaling of the maximum number of contacts}
The ``Combinatorial Kepler Problem'' asks how $M(n)$ behaves as $n\to\infty$.  We expect $M(n)\sim6n$ to leading order as the cluster approaches a close-packing, but one can also include surface corrections, which scale as $n^{2/3}$. Bezdek et al. \cite{bezdek2012,bezdek2013b} proved that $6n - 7.862n^{2/3}\leq M(n) \leq 6n - 0.926n^{2/3}$, where the upper bound holds for all $n$ and the lower bound holds for $n=6,19,\ldots,k(2k^2+1)/3$, $k \in \mathbb{N}$. We find that $M(n) =  \lceil \text{lower bound} \rceil + 1$ for $6\leq n\leq 19$, (except $n=12$ where the correction is 2), and the upper bound is more than double the maximum; this suggests the lower bound is closer to the correct scaling so the bounds could be strengthened. 
Table \ref{tbl:bezdek} shows the maximal number of contacts $M(n)$ found by our algorithm, and the lower and upper bounds for $M(n)$ as proven in \cite{bezdek2012,bezdek2013b}. The upper bound is proven for all $n$ and the lower bound for $n=k(2k^2+1)/3$, $k\in\mathbb{N}$. 

\begin{table*}[!h]
\hspace*{-10pt}\makebox[\linewidth][c]{
\begin{tabular}{| l | ccccccccccccccc}
$n$ & 5 & 6 & 7 & 8 & 9 & 10 & 11 & 12 & 13 & 14 & 15 & 16 & 17 & 18 & 19 \\\hline
ceil($6n-7.862n^{2/3}$) & 8 & 11 & 14 & 17 & 20 & 24 & 28 &31 & 35 & 39 & 43 & 47 & 51 & 55 & 59 \\
$M(n)$ & 9 & 12 & 15 & 18 & 21 & 25 & 29 & 33 & 36 & 40 & 44 & 48 &  52 & 56 & 60 \\
floor($6n-0.926n^{2/3}$) & 27 & 32 & 38 & 44 & 49 & 55 & 61 & 67 & 72 & 78 & 84 & 90 & 95 & 101 & 107 \\\hline
\end{tabular}
}
\caption{Upper and lower bounds for the combinatorial Kepler problem, and our data. The upper bound is proven for all $n$ and the lower bound for  $n=k(2k^2+1)/3$, $k\in\mathbb{N}$. }\label{tbl:bezdek}
\end{table*}

\begin{table*}
\begin{tabular}{c | cccccccc }
$n$ & \multicolumn{7}{c}{\# of close-packing fragments (total \# of clusters)} \\ 
   & $3n-7$ & $3n-8$ & $3n-7$ & $3n-6$ & $3n-5$ & $3n-4$ & $3n-3$& $3n-2$ \\ \hline
5 & & & & 1 (1)  &&&&\\
6 & & & & 1 (2)  &&&&\\
7 & & & & 1 (5)  &&&& \\
8 & & & & 4 (13)  &&&& \\
9 & & & & 11 (52)  &&&&\\
10 & & & 0 (1) & 33 (259) & 3 (3)  &&& \\  
11 & & 0 (2) & 4 (18) & 103 (1618) & 12 (20) & 1 (1)  && \\
12 & & 0 (11) & 13 (148) & 339 (11,638) & 77 (174) & 4 (8) & 1 (1) & \\
13 & & 1 (87) & 57 (1221) & 1079 (95,810) & 364 (1307) & 42 (96) & 5 (8) & \\
14 & 0 (1) & 6 (707) & 242 (10,537) & 3451 (872,992) &  1622 (10,280) &  298 (878) &  35 (79) & 2 (4) \\\hline
\end{tabular}\\

\begin{tabular}{c | cccccccc  }\hline
 & $3n-4$ & $3n-3$ & $3n-2$ & $3n-1$ & $3n$ & $3n+1$ & $3n+2$ & $3n+3$\\\hline
15 &     1748 (7675)  & 265 (782)  &  23 (55) & 5 (6) & & & &  \\
16 &       &  1997 (7895) & 220 (664) & 29 (62) & 7 (8)  &&& \\
17 &     &    & 2036 (7796) & 267 (798)  &  51 (85) & 6 (6) && \\
18  &&& & 2451 (9629) & 434 (1085) & 59 (91) & 5 (5) &  \\
19  &&& &  & 3727 (13472) & 681 (1458) & 64 (95) & 6 (7) \\ \hline
\end{tabular}\\

\begin{tabular}{c | c c}\hline
$n$ & Total close-packing fragments (Total clusters)&  \% Close-packing fragments\\\hline
5 & 1(1) & 100\% \\
6 & 1(2) & 50\%\\
7 & 1(5) & 20\%\\
8 & 4 (13) & 31\%\\
9 & 11 (52) & 21\%\\
10 & 36 (263) & 17\%\\
11 & 120 (1659) &  7.2\% \\
12 &   434 (11,980) & 3.6\%\\
13 &  1548 (98,529) & 1.6\%\\
14 & 5656 (895,478) & 0.63\%\\\hline
\end{tabular}

\begin{tabular}{cc}
\hline
$n$ & type \\\hline
10 & hcp, hcp, 2d\\
11 & hcp \\
12 & hcp \\
13 & hcp, fcc, hcp, 2d, none, rcp, none, none   \\
14 & hcp, fcc, none, none  \\
15 & hcp, hcp, fcc, hcp, fcc, none   \\
16 & hcp, hcp, fcc, hcp, hcp, fcc, fcc, none  \\
17 & hcp, hcp, hcp, fcc, fcc, hcp  \\
18 & hcp, hcp, hcp, fcp, hcp \\ 
19 & hcp, hcp, hcp, fcc, fcc, rcp, none  \\ \hline
\end{tabular}

\caption{Close-packing fragment data. Top: number of close-packing fragments, organized by number of contacts. The total number of clusters of each type is shown in brackets. Middle: total number of close-packing fragments for each $n$.
Bottom: close-packing fragment type for ground-state clusters, in the order they are shown in Figure \ref{fig:max} 
(if shown). Here ``fcc, hcp, rcp, 2d, none'' stand for face-centered cubic, hexagonal close packing, random-stacking (neither fcc nor hcp), two-dimensional lattice fragment (undetermined), and defective respectively.}\label{tbl:lattice}
\end{table*}

\begin{table*}
\begin{tabular}{c | cccccccc }
$n$ & \multicolumn{8}{c}{\# of singular clusters (total \# of clusters)} \\ 
   & $3n-9$ & $3n-8$ & $3n-7$ & $3n-6$ & $3n-5$ & $3n-4$ & $3n-3$ & $3n-2$ \\ \hline
9 && & & 1 (52)  &&&\\
10 && & 1 (1) & 4 (259) & 0 (3)  &&  \\  
11 && 2 (2) & 18 (18) & 28 (1618) & 1 (20) & 0 (1)  &   \\
12 && 11 (11) & 148 (148) & 175 (11,638) & 7 (174) & 0 (8) & 0 (1)   \\
13 && 87 (87) & 1221 (1221) & 1311 (95,810) & 50 (1307) & 1 (96) & 2 (8)    \\
14 &1(1) & 707 (707)& 10,537 (10,537)& 10,390 (872,992)& 416 (10,280)& 4 (878)& 3 (79)& 0 (4)\\\hline
\end{tabular}\\

\begin{tabular}{c | cccccccc  }\hline
 & $3n-4$ & $3n-3$ & $3n-2$ & $3n-1$ & $3n$ & $3n+1$& $3n+2$ & $3n+3$\\\hline
15  & 54 (7675)  & 14 (782)  &  0 (55) & 0 (6) & & & & \\
16   &   &  87 (7895) & 0 (664) & 0 (62) & 0 (8) & & & \\
17 & & & 10 (7796) & 0 (798)  & 0 (85) & 0 (6) & & \\
18 &&&& 13 (9629) & 0 (1085) & 0 (91)  & 0 (5) &  \\
19 &&&&  & 31 (13472) & 0 (1458)  & 0 (95) & 0 (7) \\\hline
\end{tabular}\\

\begin{tabular}{c | c c}\hline
$n$ & Total singular (Total clusters) &  \% Singular\\\hline
9 & 1 (52) & 1.9\%\\
10 & 5 (263) & 1.9\%\\
11 & 49 (1659) & 2.95\% \\
12 &   341 (11,980) & 2.85\%\\
13 &  2672 (98,529) & 2.71\% \\
14 & 22,058 (895,478) & 2.46\%\\\hline
\end{tabular}\\

\begin{tabular}{cc | c}\hline
n & close-packing\&singular (total singular) & \% \\\hline
9 & 1 (1)   & 100\% \\
10 & 2 (5)  & 40\%\\
11 & 13 (49) & 26.5\%\\
12 & 40 (341) & 11.7\%\\
13 & 174 (2672) & 6.5\% \\
14 & 791 (22,058) & 0.088\%\\\hline
\end{tabular}

\caption{Singular cluster data. Top: number of singular clusters, organized by number of contacts, with the total number of clusters of each type shown in brackets.
Middle: total number of singular clusters. Bottom: total numbers of clusters which are both close-packing fragments, and singular.}\label{tbl:singular}
\end{table*}

\clearpage


\section{Numerical Algorithm}

The numerical algorithm consists of three components: one to determine the dimension of the solution set to equation
\eqref{eq:bonds}, \eqref{eq:cons}, 
one to project onto the solution manifold, and one to walk along a one-dimensional the manifold. We describe these in turn, as well as the algorithm to store and compare clusters.  

\subsection{Determining the dimension of the solution set}\label{sec:dim}

Suppose we have an adjacency matrix $A$ representing a set of $m$ contacts, and a solution $\b{x}$ to the corresponding system of equations \eqref{eq:bonds},\eqref{eq:cons}.
There is one equation for each pair of spheres $(i,j)$ that are in contact, and the translational and rotational degrees of freedom are removed by fixing the first sphere at the origin, the second on the $x$-axis, and the third on the $xy$-plane. In practice, we choose three spheres that form a triangle of contacts, to avoid problems if the three spheres lie on a line. All clusters encountered contain at least one such triangle. We wish to determine the dimension of the solution set near $\b{x}$, assuming that $\b{x}$ lies on a smooth manifold so this quantity is well-defined. In particular, we wish to determine whether the dimension is 0, 1, or $>1$. 

We do this in two ways. 
One is a semi-analytic method that determines the dimension via solvability conditions. This can be rigorously justified, but is not developed to handle every possible case that can appear in practice. 
The second  is purely numerical, based on trying to move small distances in each of the candidate tangent directions. It produces an answer for all cases, and though it is not rigorous, it still worked extremely well in that it gave the correct dimension in cases we could verify with the semi-analytic method. However, it was orders of magnitude slower than the latter, which was necessary to apply our algorithm to larger values of $n$. 

Both methods require looking at derivatives of \eqref{eq:bonds}. Let us look for an analytic motion of the sphere centers $\b{p}(t)$ parameterized by $t\in\R$ with $\b{p}(0) = \b{x}$. Differentiating \eqref{eq:bonds} gives
\begin{align}
R(\b{x})\b{p'} &= 0 \label{eq:order1}\\
R(\b{x})\b{p''} &= -R(\b{p'})\b{p'} \label{eq:order2}\\
\vdots\qquad & \qquad\quad\vdots \nonumber\\
R(\b{x})\b{p}^{(n+1)} & -\sum_{k=1}^n \binom{n}{k} R(\b{p}^{(k)})\b{p}^{(n-k+1)} \label{eq:ordern}
\end{align}
where $R(\b{p})$ is the rigidity matrix. It is constructed so that $(R(\b{p})\b{y})_k = (\b{p}_{i_k} - \b{p}_{j_k})\cdot ( \b{y}_{i_k} - \b{y}_{j_k})$ for any vector $\b{y}\in\R^{3n}$ and each contact $k$, $1\leq k \leq m $. From the final six equations we have $R_{m+j} = e_{s_j}^T$ for $1\leq j\leq 6$, $s_j\in\{1,2,3,5,6,9\}$ for the constrained vertices, where $e_s$ is the vector with 1 in the $s$th position and zeros elsewhere. 
Let the right and left null spaces of the rigidity matrix be $\mathcal{V}$, $\mathcal{W}$ respectively, with bases $\{v_i\}$,   $\{w_i\}$ and sizes  $n_v = |\mathcal{V}|$, $n_w = |\mathcal{W}|$.

If $R(\b{x})$ is of full rank and the number of rows does not exceed the number of columns, then the solution set is regular, the Implicit Function Theorem applies, and the dimension of the solution set is $n_v$. Otherwise, the solution set is singular, and we use the following numerical algorithms to determine its dimension.

\paragraph{Semi-analytic determination of dimension}
Let the tangent space to the solution to \eqref{eq:bonds} have orthogonal basis $\mathcal{B}$, and let $D=|\mathcal{B}|$ be its dimension. This method proceeds as follows: 
\begin{enumerate}
\item If $|\mathcal{V}|=0$: the cluster is \emph{first-order rigid}. Return $D=0$. 
\item If $|\mathcal{W}|=0$: there are no solvability conditions on \eqref{eq:order1}--\eqref{eq:ordern}, so this system can be solved up to any order. Proceed to numerical method, or return $D = n_v$. 
\item Test for second-order rigidity. To solve \eqref{eq:order2} for $\b{p}'\in \mathcal{V}'$, where $\mathcal{V}'\subset\mathcal{V}$ is a linear subspace, the Fredholm Alternative requires that $w^TR(v)v=0$ for all $w\in\mathcal{W}$, $v\in\mathcal{V}'$. When this does not hold for any subspace $\mathcal{V}'$, the cluster is second-order rigid \cite{connelly1996}.

An arbitrary right and left null vector can be written as  $v = \sum_j a_j v_j$, $w=\sum_j b_j w_j$, with $a = (a_1,\ldots,a_{n_v})$, $b=(b_1,\ldots,b_{n_w})$. The RHS of \eqref{eq:order2} can be written as $-\sum_{i,j} a_ia_j R(v_i)v_j$. Taking the inner product with $w$ yields the following:
\begin{equation}\label{eq:Q}
a^T\left( \sum b_k Q^{(k)}\right)a = 0, \quad Q^{(k)}_{ij} = w_k^TR(v_i)v_j . 
\end{equation}
If we can find a linear subspace of $a$-values such that this holds for all $b\in\R^{n_w}$, then we can solve \eqref{eq:order2}, and the tangent space is contained in the space $\mathcal{V}'$ spanned by $\sum a_j v_j$. 
The negation requires showing that for every $a\in\R^{n_v}$, there is a $b\in\R^{n_w}$ such that \eqref{eq:Q} is non-zero, and then the cluster is second-order rigid. 
This is hard to show in general, but what is possible is to find a $b$ such that $ \sum b_k Q^{(k)}$ is sign-definite. The left null vector $w=\sum_j b_j w_j$ ``blocks'' all right null vectors and the cluster is called \emph{pre-stress stable}. It is possible to find such a $b$ using semi-definite programming (SDP) methods, for example. 

In practice, we only check the vectors $b_k=e_k$. If any matrix formed from these is sign-definite, return $D=0$. Otherwise,  continue to the next step. 
We do this because we got lucky: this test always agreed with our numerical algorithm
(except for 7 clusters at $n=14$ that we left out of the dataset.) 
Enumerating larger $n$ where unusual cases are more likely to occur will require implementing SDP methods.


\item If the dimension is still undetermined, continue to the numerical method. 

\end{enumerate}

 It was shown in \cite{connelly1996} that if  the cluster is either first-order-rigid or second-order rigid as described in the tests (1., 3.) above, then it is rigid, so these tests are sufficient to prove rigidity (up to numerical tolerance.) Unfortunately there is no equivalent notion of higher-order rigidity, because it could be that every analytic parameterization of cluster motion has $\b{p}'(0)=0$, i.e. the cluster corresponds to a ``cusp,'' in which case there is a different system of equations to solve \cite{connelly1994,garcea2005}. 
It would be useful to extend these solvability conditions to higher orders and cusps, even if they do not prove rigidity, because this would still provide useful information about a cluster's stability. 
 

\paragraph{Numerical determination of dimension}
This method tries to estimate an orthogonal basis $\mathcal{B}$ for the tangent space by doing the following for each $v_j\in\mathcal{V}$:
\begin{enumerate*}
    \item Take a step of size $\Delta s_0$ in directions $\pm v_j$ to obtain $x_\pm = \b{x}\pm \Delta s_0\: v_j$
    \item Project back onto the manifold of constraints to obtain $x_\pm' = \text{Proj}\left(x_\pm\right)$. Initially we also require $(x_\pm'-x_\pm)\perp v_j$ to prevent the projection from going back to the starting point, but if this fails we relax the condition.  
    \item Let $v_t = x_\pm'  - \b{x}$ be the estimated tangent vector. If $|v_t| > \texttt{xTolMax}$ or $|v_t| < \texttt{xTolMin}$, reject the vector. Otherwise, project $v_t$ onto our current estimate of $\mathcal{B}$, and 
    let $v_t^\perp / |v_t|$ be orthogonal to the projection. 
    \item If $|v_t^\perp| > \texttt{vTol}$, then add $v_t^\perp$ to the basis $\mathcal{B}$. 
\end{enumerate*}

\bigskip

We use both methods to determine whether to follow a path or not, but only use the semi-analytic method to decide whether or not a cluster is rigid. Therefore all clusters we list are pre-stress stable up to numerical tolerance.

\subsection{Projecting onto a manifold}

If we have an approximate solution $\b{x'}$ to \eqref{eq:bonds}, we obtain a more accurate solution by solving \eqref{eq:bonds} using Newton's method with a given tolerance \texttt{Tol}. This is not an orthogonal projection, but we compared it with the orthogonal projection described in \cite{holmescerfon2013} which works for regular clusters, and found the two to be very close.
We imposed a maximum step size of $\Delta x_{max}$ in each Newton's iteration to avoid moving too far away from the solution. 

This method suffers several drawbacks for singular clusters. First, Newton's method loses its quadratic order of convergence, so it typically took an order of magnitude more iterations to converge, and occasionally it never converged (this happened only very rarely -- there were 1548 total projection errors of all types for $n=13$.) Second, even though the constraints are satisfied to tolerance \texttt{Tol}, the actual solution is less accurate if it is singular. For example, if the equation $x^2=0$ is satisfied to order $\epsilon$, then we expect $x\approx \sqrt{\eps}$. 
We tested the accuracy of the cluster coordinates by perturbing each cluster by some large amount and re-projecting, and found that all clusters tested (including the very hypostatic ones) appeared to be accurate to within $\sqrt{\texttt{Tol}}$. In general, one would expect to lose accuracy as the nature of the singularity worsens.

These concerns can be mitigated through the use of deflation techniques, see e.g. \cite{dayton2005, lvz2006, dayton2011, wampler2011}. 
We tried these, but found them not as useful for our study because they require doubling the number of variables at each deflation step. For many singular clusters we had to apply several deflation steps before we obtained quadratic convergence and linear accuracies, however for these clusters we achieved accuracies of $\sqrt{\texttt{Tol}}$ anyways so the huge slowdown due to the extra variables was not worth the computational effort. 

\subsection{Moving along a one-dimensional manifold}

Once we determine that a solution set to \eqref{eq:bonds} is one-dimensional, we move along it as follows: first, we extract the direction(s) that make the broken contacts increase in length. For each direction, we alternate between taking a step of size $\Delta s$ along the manifold in the tangent direction $v_k$, and  projecting back onto the manifold. After each projection we form the rigidity matrix in \eqref{eq:order1}, compute its null space $\mathcal{V}$, and find the next tangent direction $v_{k+1}$ by the least-squares projection of $v_k$ onto $\mathcal{V}$. This ensures that we keep going in the same direction, i.e. we don't accidentally start moving backwards along the manifold, and it provides an estimate of the single tangent direction when the path is singular. 

After the first step, we check the dimension as in section \ref{sec:dim}, and stop moving if this dimension has increased or decreased. 
For $n=13$  it increased for 3851 paths and decreased for 23. 
We do not check the dimension after the first step, as this is very time-consuming. 

At each step we check whether two spheres initially not in contact are within some tolerance $1+\texttt{tolA0}$. 
The first time this happens we back up one step and repeat the continuation with a smaller step size $\Delta s_0$, and again stop when spheres are within $1+\texttt{tolA0}$.
Then, we project onto this new set of constraints and check if the cluster is rigid, using a new tolerance \texttt{tolA} to determine whether two spheres are adjacent. If, as happens very occasionally, this projection fails (for example because \texttt{tolA0} is deliberately chosen too big initially, to ``catch'' more pairs than are actually adjacent), we delete subsets of the new constraints until the projection succeeds. If it never succeeds we abandon the cluster. Note that $\texttt{tolA0}$ should be chosen comparable in magnitude to $\Delta s$ because sometimes spheres can come into contact tangentially.

\subsection{Cluster isomorphism}

We keep track of clusters through their adjacency matrices in a hash table. Each adjacency matrix has a canonical form that we compute using \texttt{nauty} \cite{mckay1981}. This is converted to a binary vector $a \in \{0,1\}^{n^2}$  and we define an ordering by setting $a<b$ if $a_k < b_k$  where $k$ is the first entry where they differ. 

Adjacency vectors are stored as a binary tree. Each leaf contains the indices of the clusters with that adjacency matrix, and the indices of the child adjacency vectors that are larger or smaller. To add an adjacency vector $a$ we compare it to a leaf $b$ on the tree, and move to the larger or smaller child depending on whether $a>b$ or $a<b$. We add $a$ as a child to the leaf at the end of the tree. 

When we find a new cluster, we compute the canonical form of the adjacency matrix and coordinates. If this adjacency matrix is already in the list, we compare the cluster and its reflection to those with the same adjacency matrices, by rotating so the same spheres on each are at the vertices of a given equilateral triangle. We use a tolerance \texttt{tolD} to determine if the coordinates are the same.

\subsection{Numerical parameters}

Here are the values of the numerical parameters used in most of the simulations. 
For $n=$11--13 we ran several simulations with different choices of parameters, and combined data if necessary. 
Typically the datasets for $n=12,13$ differed by an $O(1)$ number of clusters, while for $n=11$ they were typically identical.
All numerical computations were performed in double precision. \\

\begin{tabular}{llp{6cm}}
Parameter & Value & Description \\\hline
$\Delta s$ & 5e-3 & step size along path, in endgame\\
$\Delta s_0$ & 5e-2 & step size along path, initially\\
\texttt{Tol} & $9e-16$ & tolerance for Newton's method when projecting / sharpening cluster coordinates\\
\texttt{TolN} & 1e-6 & tolerance on singular values for null space of rigidity matrix \\
$\Delta x_{max}$ & 0.02 & maximum step size in Newton's method\\
\texttt{tolA} & 1e-5 & tolerance for spheres being adjacent \\
\texttt{tolA0} & 1e-3 & initial tolerance for spheres being adjacent, used to stop following path \\
\texttt{tolD} & 1e-5 & tolerance for coordinates of a cluster being identical\\
\texttt{xTolMax} & 10$\Delta s$ & upper bound on cluster displacement, to determine if a cluster has moved along tangent space \\
\texttt{xTolMin} & $\Delta s$/8 & lower bound on cluster displacement, to determine if a cluster has moved along tangent space \\
\texttt{vTol} & 2$\Delta s$ & tolerance for determining whether a unit vector is orthogonal to the estimated tangent space: the part orthogonal to the projection must have at least this magnitude. Depends on step size $\Delta s$ used to move in tangent space, because of curvature of manifold. \\
\end{tabular}\\

\bigskip

The value of \texttt{tolA} was chosen to stay well away from the resolution of $\sqrt{\texttt{Tol}}\approx 3e-8$ of the coordinates of the singular clusters, to avoid computing junk clusters. 
For $n\geq 15$ it will be unable to resolve the gaps of certain regular, non-hyperstatic clusters. These gaps may be resolvable in double precision for a few more values of $n$  because \texttt{tolA}  can be increased and still be larger than the numerical precision of the cluster coordinates. However, as $n$ increases further the distances between non-contacting spheres are expected to become arbitrarily close to 1, so higher precision will be necessary.

\end{document}